\documentclass[%
 reprint,
superscriptaddress,
showpacs,
 amsmath,amssymb,
 aps,
 pra,
]{revtex4-1}

\usepackage{graphicx}
\usepackage{dcolumn}
\usepackage{bm}
\usepackage{placeins}
\usepackage{color}

\begin{document}

\preprint{APS/123-QED}

\title{Disentangling Structural Information From Core-level Excitation Spectra}

\author{Johannes Niskanen}
\email{johannes.niskanen@helmholtz-berlin.de}
\affiliation{Helmholtz Zentrum Berlin f\"ur Materialien und Energie, Institute for Methods and Instrumentation for Synchrotron Radiation Research, Albert-Einstein-Str. 15, D-12489 Berlin, Germany}

\author{Christoph J. Sahle}
\affiliation{ESRF-The European Synchrotron, 71 Avenue des Martyrs, F-38000 Grenoble, France}%

\author{Keith Gilmore}
\affiliation{ESRF-The European Synchrotron, 71 Avenue des Martyrs, F-38000 Grenoble, France}

\author{Frank Uhlig}
\affiliation{Universit\"at Stuttgart, Institute for Computational Physics, Allmandring 3 D-70569 Stuttgart,
Germany}

\author{Jens Smiatek}
\affiliation{Universit\"at Stuttgart, Institute for Computational Physics, Allmandring 3 D-70569 Stuttgart,
Germany}

\author{Alexander F\"ohlisch}
\affiliation{Helmholtz Zentrum Berlin f\"ur Materialien und Energie, Institute for Methods and Instrumentation for Synchrotron Radiation Research, Albert-Einstein-Str. 15, D-12489 Berlin, Germany}
\affiliation{Universit\"at Potsdam, Institut f\"ur Physik und Astronomie, Karl-Liebknecht-Strasse 24/25 D-14476 Potsdam-Golm, Germany} 

\date{\today}

\begin{abstract}
Core-level spectra of liquids can be difficult to interpret due to the presence of a range of local environments. We present computational methods for investigating core-level spectra based on the idea that both local structural parameters and the X-ray spectra behave as functions of the local atomic configuration around the absorbing site. We identify correlations between structural parameters and spectral intensities in defined regions of interest, using the oxygen K-edge excitation spectrum of liquid water as a test case. Our results show that this kind of analysis can find the main structure--spectral relationships of ice, liquid water, and supercritical water.
\end{abstract}

\maketitle

\section{Introduction}
At given thermodynamic conditions, atomistic systems occupy points $\bf R$ in the configurational space with probabilistic weights $\rho ({\bf R})$ that depend on the statistical ensemble \cite{tuckermanbook}. Determining these probabilistic weights as a function of the thermodynamic conditions is of considerable interest. Core-level spectroscopy is sensitive to local atomic and electronic structure, in part, because the transition matrix elements depend on the spatial overlap of the strongly localized core level with the valence levels. Because distant atoms typically have only a weak effect on core spectra \cite{niskanenGly} the most significant effects on spectral shape originate from the closest neighbors surrounding the excited site. These nearest neighbors, together with the excited atom, define a local configuration. One expects to find strong correlations between this local configuration, classified in terms of selected structural parameters, and the excitation spectrum. Developing methodology to identify these correlations is the subject of this work.
\par
Electronic transitions are not always as easy to interpret as to measure. A typical analysis technique of experimental X-ray spectra involves fitting procedures for peak positions and intensities, which can be used to identify the underlying structure. Such analysis of experimental spectra is complicated because it only deals with the statistical average spectrum, and the underlying individual atomistic structures may have considerable variation in their core-level spectra. On the other hand, sometimes appreciable changes in atomic positions cause only subtle spectral changes \cite{niskanenJCP}. Thus, not all structural information is manifested in the spectra of the system, which raises a deeper question of how to find the structural degrees of freedom that the experiment is sensitive to.
\par
To gain more insight into the experiment, calculations may be performed to interpret spectral lines and their connection to structural degrees of freedom. For crystalline materials, existing structural knowledge and geometry optimization can be used to obtain structures for spectral evaluation. This common approach is computationally light and still allows interpretation of the data in a meaningful fashion \cite{inkinen}. Liquid systems, on the other hand, exhibit a continuous range of configurations. They explore the accessible phase space and, therefore, computational studies of liquids require statistical sampling \cite{allentildesley}. This sampling can be accomplished effectively through molecular dynamics simulations, which is the approach we take here.
\par
In this work, we study correlations between structural parameters of the local environment of the absorbing site and intensities in regions of interest (ROI) in the corresponding core-level excitation spectra. Statistical sampling of atomic structures is done by molecular dynamics while spectra are obtained from first-principles calculation. We use liquid water as the benchmark for this procedure, with future studies of systems with a large number of degrees of freedom in mind. Compared to liquid water, ice has stronger tetrahedral structure and hydrogen bonding, and increased intensity in the post-edge region \cite{tse2008} of the O K-edge excitation spectrum. Supercritical water, on the other hand, has more broken hydrogen bonds and greater deviation from the lattice-like structure with prominent pre-edge feature in the O K-edge spectrum \cite{sahlePNAS}. The two analysis methods used, classification and linear correlation coefficients, find these structure--spectrum relationships in a single simulation of liquid water.

\section{Model and Methods}\label{modelandmethods}
The nearest neighbors of the excited atomic site (or lack of them) typically cause the most significant effects on its X-ray spectrum. Including the excited atom, we define this atomic cluster as the local configuration ${\bf R}^{(loc)}$. The thermodynamic distribution causes an ensemble of local configurations, which also yields a distribution of core-level spectra. The challenge is to extract meaningful information by relating aspects of these local structures to intensities of spectral features. First, we define regions of interest (ROI) in the simulated ensemble average spectrum, which is followed by integration of intensity from each local configuration over this ROI. This procedure allows us to use a single number, $F({\bf R}^{(loc)})$, to quantify the intensity contribution of a local configuration to a particular spectral feature. Second, we define structural parameters, such as the number of hydrogen bonds of the probed atom, as functions of the local configuration. In general, such a parameter is a real-valued quantity $P({\bf R}^{(loc)})$ for which some mathematical definition needs to be chosen. We therefore relate a local configuration to both its structural parameters and the ROI intensities of the resulting spectrum. While $F$ and $P$ do not generally have a 1-to-1 correspondence, statistical correlations between these quantities can be obtained, which give insight in how line intensities are related to underlying structures.

\subsection{Computational Methods}
Atomic structural models of water were generated by {\it ab initio} molecular dynamics (AIMD), as described in detail in Ref. \citenum{frank}. The system of 64 water molecules was treated in (1.24 nm)$^3$ cubic cell with periodic boundary conditions. The system was first equilibrated in the canonical (NVT) ensemble at 330 K for about 20 ps using a velocity rescaling thermostat \cite{bussi2007}, after which the dynamics was continued in the microcanonical (NVE) ensemble for about 200 ps. The resulting temperature in the NVE simulation was around 340 K. Accounting for the quantum dynamics of the hydrogen nuclei, such as via path integral molecular dynamics (PIMD), gives improved radial distribution functions for water, which then yields spectra with pre-edge features in closer agreement with experiment.  However, PIMD simulations are considerably more intensive then AIMD.  As a compromise, it has been found that increasing the simulation temperature to 330 K in AIMD produces results for the O-O and O-H radial distribution functions in close agreement to 300 K results from PIMD simulations \cite{watermorrone}.
\par
\begin{figure}[h]
\begin{center}
\includegraphics[width=8.75cm]{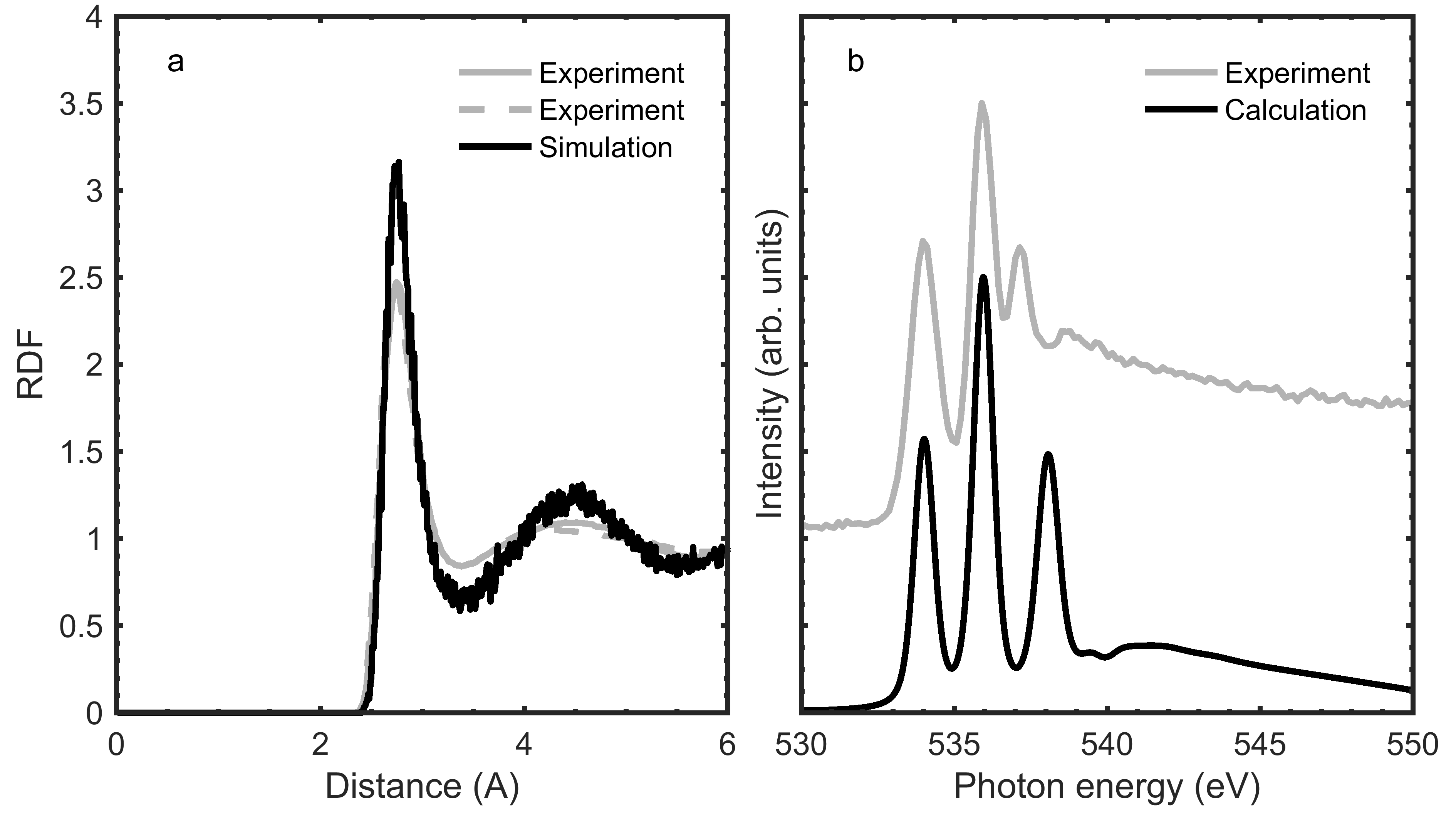}
\end{center}
\caption{\label{gasrdf} The O-O radial distribution function from the AIMD simulation compared to experiment (Ref. \citenum{soper2007}) (a). The calculated and experimental (Ref. \citenum{hitchcockwater}) O1s excitation spectra of gas phase water (b). The simulated spectrum was shifted to be aligned at the lowest resonance.}
\end{figure}

Forces for the integration of Newton's equation of motion were obtained using density functional theory as implemented in the Quickstep \cite{vandevondele2014} module of CP2K \cite{hutter2014}. We used the BLYP exchange-correlation functional \cite{becke1993,lee1988} together with a pairwise-additive dispersion correction \cite{grimme2006}. The Kohn-Sham orbitals were expanded into a combined plane-wave and Gaussian orbital basis set, with a kinetic energy cutoff of 400 Ry and a dual-$\zeta$ representation \cite{vandevondele2007}, respectively. Core-electrons were replaced by separable dual-space Gaussian pseudopotentials \cite{PP}. The O-O radial distribution function from the simulation is shown in Figure \ref{gasrdf} (a) in comparison to experiment from Ref. \citenum{soper2007}.
\par
Individual non-resonant inelastic X-ray scattering spectra were calculated for each oxygen atom of 21 snapshots sampled along the production run trajectory by using the Bethe--Salpeter equation (BSE) method as implemented in the OCEAN code \cite{bse,gilmore2015}. It has previously been demonstrated that the BSE method, as implemented in the OCEAN code, can produce satisfactory X-ray absorption spectra of liquid and ice-phase H$_2$O \cite{vinsonice}. For further reference, in Figure \ref{gasrdf} (b) we compare the calculated spectrum of a single geometry-optimized H$_2$O molecule with the experimental gas-phase spectrum taken from Ref. \citenum{hitchcockwater}. The sampling procedure of the AIMD simulation yields 1344 spectra, each of which holds unique information about the local chemical and structural environment ${\bf R}_i^{(loc)}$ around the respective excited oxygen atom ($i=1,...,1344$). Averaging over all individual oxygen K-edge spectra yields the average spectrum (solid black line in Fig.~\ref{spectrabse}), which can be compared to the experimental data (solid grey line in Fig.~\ref{spectrabse}). We used a momentum transfer of 3.1 \AA$^{-1}$ for the simulated spectra and convoluted all individual spectra with a GW self-energy obtained from a many-pole approximation to the valence level loss function \cite{kas2007} and a Gaussian of 0.6 eV full width at half maximum (FWHM). These values are typical for experimental non-resonant inelastic X-ray scattering spectroscopy data recorded using the large solid angle spectrometer at the inelastic X-ray scattering beamline ID20 of the European Synchrotron Radiation Facility (ESRF). These are also the values for the presented experimental spectrum of liquid water.
\par
Since OCEAN uses pseudopotentials and yields spectra with respect to the Fermi level of each snapshot, the spectra from different MD simulation snapshots are not necessarily on the same absolute energy scale. To calibrate spectra from snapshot to snapshot, the total energy shift was obtained for the average spectrum of each snapshot such that the average snapshot shows the pre-peak at 535 eV. This rigid shift was then applied to all 64 individual spectra from the corresponding snapshot. The simulated gas phase spectrum of Figure \ref{gasrdf} (b) was likewise aligned to the first resonance of the experiment. 

\subsection{Regions of Interest}
As can be seen from Figure \ref{gasrdf} (b), the X-ray spectrum of isolated H$_2$O consists of three main features that can be attributed to transitions to anti-bonding molecular orbitals of $4a_{1}$, $2b_{2}$ and $3b_{2}$ symmetry, in order of increasing energy.  The symmetries of these orbitals are largely maintained in the condensed phase \cite{watercar}, but are referred to less specifically as the pre-edge, main-edge and post-edge regions. We use these established features to naturally define three regions of interest (ROIs), specifically I=[533, 535.5], II=[535.5, 539.5], III=[539.5, 545] in eV. We assign a single numeric value for each region of interest to each absorption site by integrating the normalized spectral intensity over the energy range of the region. We label these ROI intensities as $F_{X}$ for X=I,II,III.

\subsection{Structural Characterization}
Our objective is to identify correlations between certain structural motifs and changes in the intensities of each ROI.  From the molecular dynamics simulation, we extracted 21 statistically independent structures, each consisting of 64 water molecules. This gives us a sample of 1344 unique oxygen environments. In the remainder of this section we define a set of structural parameters based on chemical intuition or common definitions from literature. The next section quantifies the correlations between the values of these structural parameters and the ROI intensities.  The basic assumption of this procedure is that spectral intensities are largely determined by the local atomic structure.  Thus, we consider only the configuration of the atoms within the first two solvation shells. The number of molecules in different solvation shells is based on oxygen-oxygen distance. The first solvation shell of the central oxygen consists of water molecules with the oxygen atom closer than 3.4~{\AA} to the central oxygen. The second solvation shell consists of molecules with the oxygen atom in the range from 3.4~{\AA} to 5.5~{\AA} from the central oxygen. These radial cutoffs are based on the minima found in the radial distribution function of Figure \ref{gasrdf}(a) after smoothing.

\subsubsection{Deviation from tetrahedrality} The angular deviation from tetrahedrality of the first solvation shell was defined based on the summed angular deviation from an angle of 109.5 degrees. We define $\mathbf{r}_i^{loc}$ as the positions of the four nearest neighbor oxygen atoms (i=1,2,3,4), measured from the excited oxygen. The angular deviation from tetrahedrality, $\Delta_a$, was defined and evaluated as
\begin{equation}
\Delta_a = \sum_{i=1}^{3}\sum_{j=i+1}^{4}|\mathrm{acos}{(\hat{\mathbf{r}}_i^{loc}\cdot\hat{\mathbf{r}}_j^{loc})}-109.5^\circ|,\nonumber
\end{equation}
where $\hat{\mathbf{r}}_k^{loc}=\mathbf{r}_k^{loc}/|\mathbf{r}_k^{loc}|$. Our definition for distance deviation from tetrahedrality, $\Delta_d$, evaluates the deviation of furthest and the closest of the four closest neighbors
$$\Delta_d=\max\{|\mathbf{r}_i^{loc}|\}_i-\min\{|\mathbf{r}_i^{loc}|\}_i.
$$
Parameters $\Delta_a$ and $\Delta_d$ are both non-negative, and apart from rounding errors (in 109.5) equal to zero for a perfect tetrahedron of any size.

\subsubsection{Hydrogen bonds}
Hydrogen bonding was evaluated for both accepted and donated bonds separately, using the criteria described in Refs.~\citenum{luzar,watercar}. To classify as hydrogen-bonded within the first coordination shell, the oxygen atoms of the two water molecules needed to be separated by less than 3.5 {\AA} and the hydrogen--donor--acceptor angle needed to be less than or equal to 30$^\circ$. For the absorbing oxygen site, we quantify the number of donating hydrogen bonds ($\#$don), the number of accepting hydrogen bonds ($\#$acc) and the total number of hydrogen bonds ($\#$tot).

\subsubsection{Additional Structural Parameters}
In addition to the five structural parameters just defined -- the number of donated, accepted, or total hydrogen bonds (\#don, \#acc, \#tot), and parameters measuring the deviation from tetrahedrality for the four nearest water molecules ($\Delta_a$ and $\Delta_d$) -- we define six additional parameters. These are the bending angle of the water molecule ($\phi$), the shorter and longer of the OH bonds ($d_{OH}^{s}$ and $d_{OH}^{l}$), and the number of water molecules in the first solvation shell (SS1), the second solvation shell (SS2), or in either shell (SS12). The distribution of all evaluated structural parameters and intensities in ROIs I, II, and III are presented in the Supplemental Material.

\subsection{Correlation Analysis}
\subsubsection{Mean-based classification}
Classification was performed by first calculating the mean of all intensities in regions I, II, and III. For each interval, each local structure was classified as above-the-average or below-the-average in intensity. The mean of each structural parameter was then evaluated for both above-the-average and below-the-average sets, again for each line region I, II, and III. Finally the difference between the two was evaluated. Error estimation was based on the 10000-fold bootstrap algorithm where the original data set was re-sampled to data of the size of the original. The analysis procedure was repeated and the standard deviation of the 10000 results for each value is given as error estimate.

\subsubsection{Linear correlation coefficients}
To search for correlated behaviour of ROI-intensity and structural parameter we calculated linear correlation coefficients between the two, for each possible ROI--parameter combination. The linear correlation coefficients measure the strength of linear dependence between the ROI intensity and structural parameter value. Therefore, ROI-intensity -- structural parameter linear correlation coefficients represent how the intensities in the different ROIs vary with the underlying structural parameters. A positive (negative) sign of the coefficient means an increase (decrease) of the ROI intensity when the parameter value increases. Before the procedure, the mean value was subtracted from each parameter set. Again, error estimation was based on the 10000-fold bootstrap algorithm where the original data set was re-sampled to data of the same size. The procedure was repeated and the standard deviation of each value is given as error estimate.

\section{Results} 
\begin{figure}[h]
\begin{center}
\includegraphics[width=8.75cm]{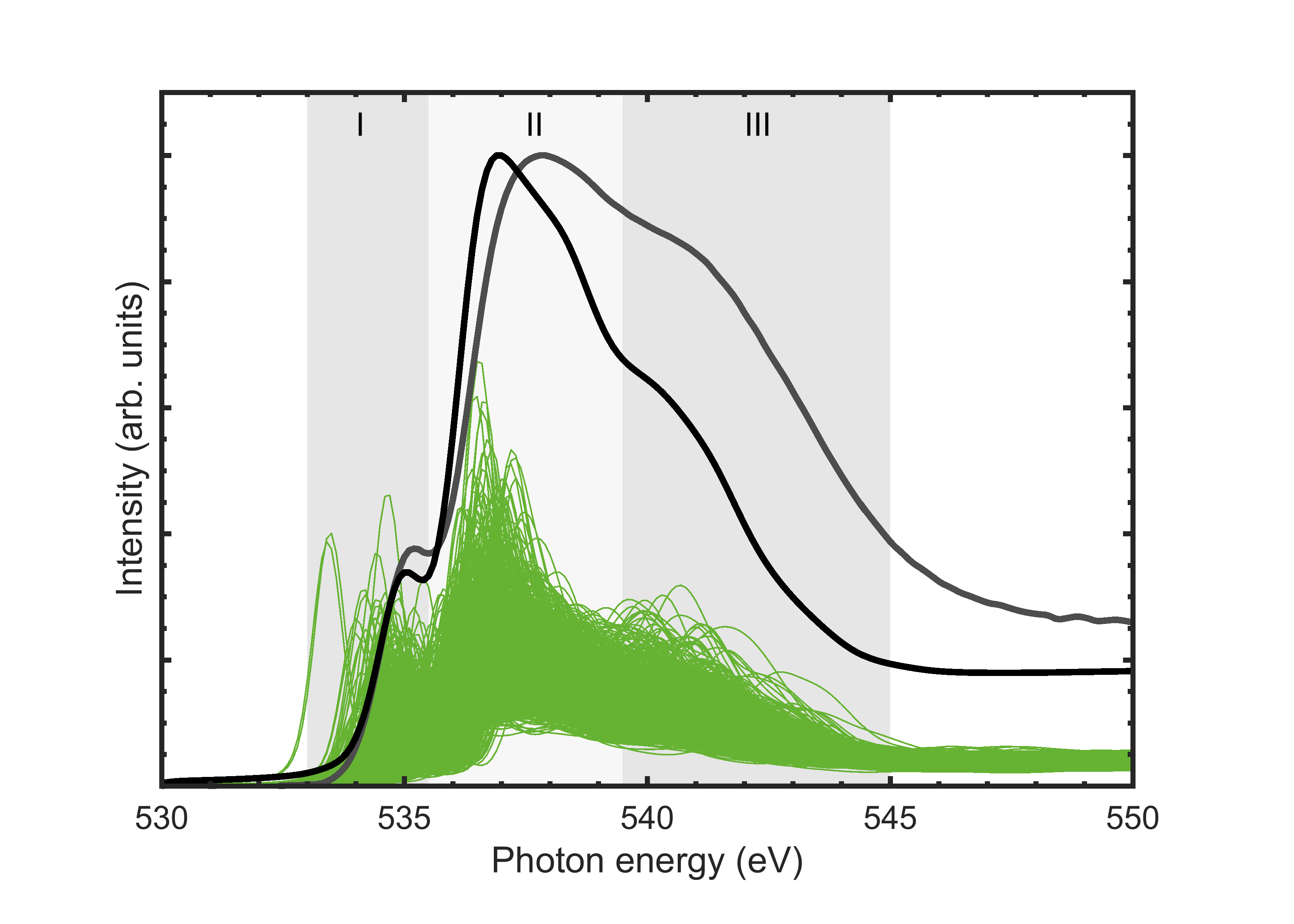}
\end{center}
\caption{\label{spectrabse} The spectra of the 1344 individual sites (green), their mean scaled by 5 (black), and the experiment (grey) on top of the shaded energy regions of interest I-III. The ROI selection is based on identifying the pre-edge, main edge and post-edge in the simulated ensemble average.}
\end{figure}

The simulated oxygen K-edge spectra of liquid water are presented in Figure \ref{spectrabse}, which presents the individual spectra as overlaying green curves with notable statistical variation. The figure shows agreement of the ensemble averaged spectrum (solid black line) with the experiment (solid gray line). All the characteristic features of the experiment are identifiably reproduced; the pre-edge, main-edge and post-edge are clearly resolved, allowing corresponding ROI selection, shown as shaded areas in the background. We assign ROI I as pre-edge, ROI II as main-edge, and ROI III as post-edge.
\par

\par
\begin{figure}[h]
\begin{center}
\includegraphics[width=8.75cm]{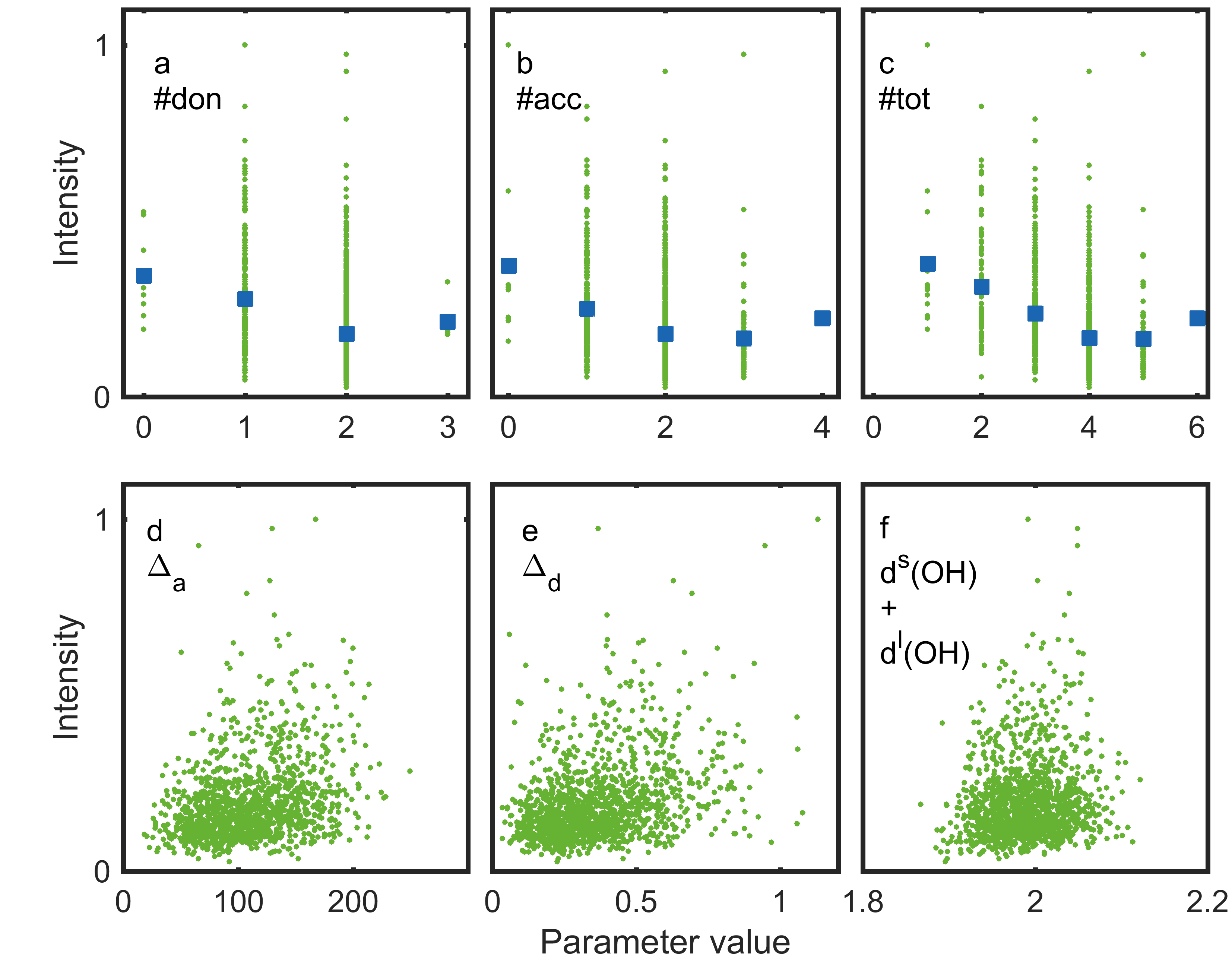}
\end{center}
\caption{\label{tetrastretchhbondbse}The integrated pre-edge (ROI I) intensity of the 1344 individual sites depicted as a function of chosen structural parameters: donated, accepted and total hydrogen bonds (a)-(c), $\Delta_a$ and $\Delta_d$ (d)-(e), and a generated parameter $d_{OH}^{s}+d_{OH}^{l}$ (f). For the hydrogen bonding parameters, the average is shown in blue.}
\end{figure}
\par
As an example of the correlation procedure, Figure \ref{tetrastretchhbondbse} presents the intensity distribution in the pre-edge (ROI I) using the hydrogen-bonding parameters in parts (a)-(c), the deviation-from-tetrahedrality parameters $\Delta_a$ and $\Delta_d$ in parts (d)-(e), and a generated parameter $d_{OH}^{s}+d_{OH}^{l}$ in part (f). The parameters related to hydrogen bonds show an increase in pre-edge intensity when donated or accepted hydrogen bonds are broken. Also, deviation from a tetrahedral geometry causes similar effects. 

We now turn to the quantified intensity -- local structure parameter interpretation. We approach the subject using two methods: mean-based classification and linear correlation coefficients. We underline that the procedure is aimed for the general case, and that liquid water is chosen because it provides the best-studied benchmark. Our motivation here is to study two simple methods, which naturally means that the list is not complete.

\subsubsection{Mean-based classification}
An example of a simple classification of structural parameter-ROI intensity ($P$-$F$) relation is based on whether a marginal increase of the weight of a local configuration in the statistical distribution increases or decreases the intensity in a given ROI. We performed this classification by first comparing intensities in ROIs I, II and III for each of the 1344 spectra to the corresponding ROI intensity of the average spectrum, and then identifying which configurations boost or suppress intensity in each ROI. Difference between the average values for the structural parameters of the two sets are presented in Figure \ref{classificationbse} normalized to the mean values. A table of the absolute difference is given in the Supplemental Material. A positive value means that larger parameter values result in an increase of ROI intensity. The error estimates represent standard deviations of the parameters in a bootstrap procedure of 10000 re-samplings of the data.
\par
The data from Figure \ref{classificationbse} reveal how different ROIs are sensitive to distinct structural motifs. For example, we can observe a correlation between the intensity in the pre-edge region (ROI I) and the deviation-from-tetrahedral parameters: the configurations showing above-average intensity in the pre-edge region have higher average parameters $\Delta_a$ and $\Delta_d$, which is manifested as positive corresponding values in Figure \ref{classificationbse}. Contrariwise, intensity in the post-edge ROI III decrease with increasing deviation from tetrahedrality. This is in line with the commonly accepted qualitative interpretation scheme of water: tetrahedrally ordered ice phases show prominent post-edge features \cite{tse2008}, whereas water in the liquid and especially in the supercritical regime exhibits very prominent pre-edges \cite{sahlePNAS}. Similarly, the increasing numbers of hydrogen bonds show decrease (increase) of the pre-edge (post-edge) in line with data from supercritical water and ice.

\begin{figure}[h]
\begin{center}
\includegraphics[width=9cm]{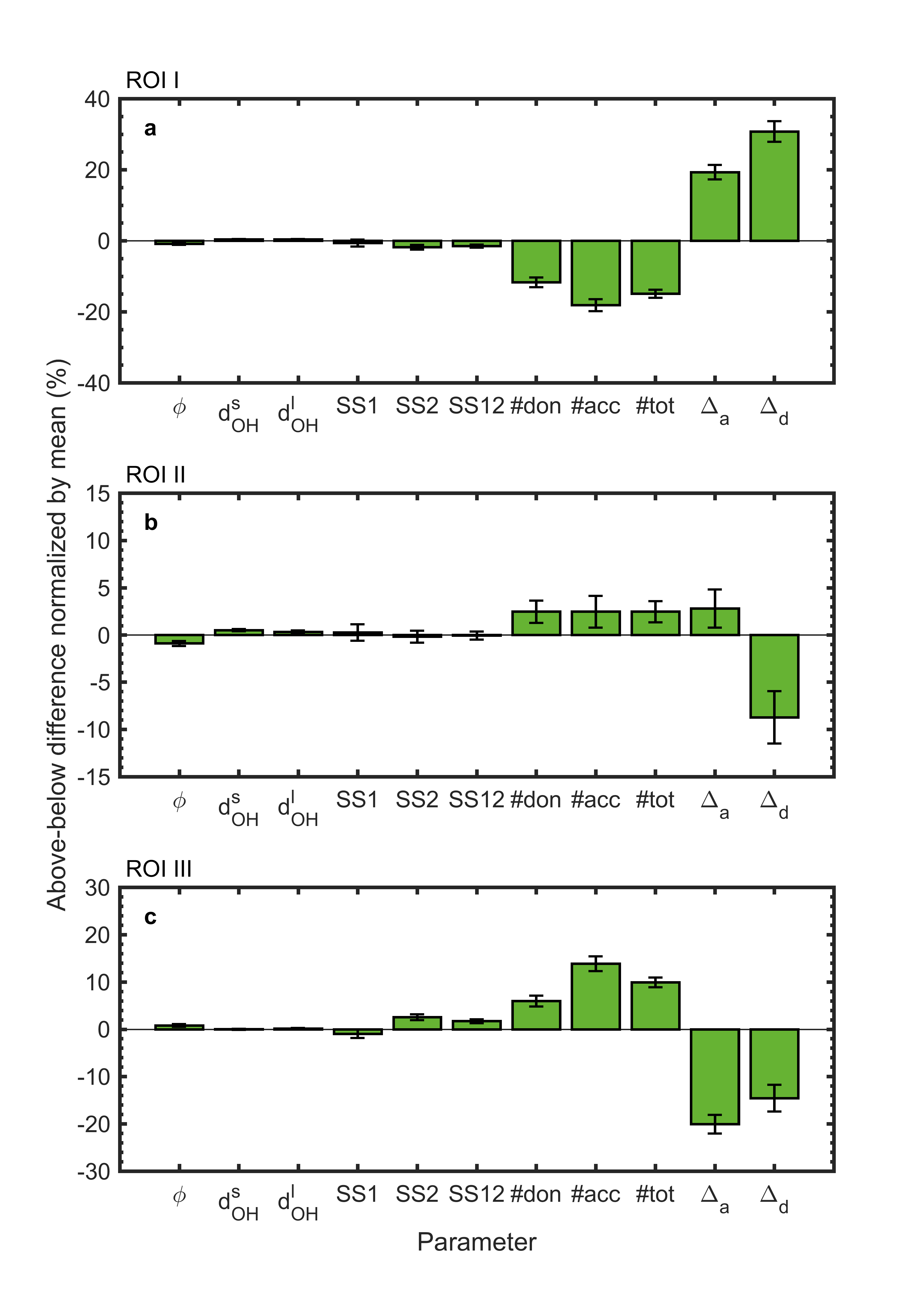}
\caption{Difference between mean structural parameters for structures boosting and suppressing average ROI intensity. Positive values mean that an increase in the parameter is associated with an increase in ROI intensity. The absolute values were normalized by the mean.\label{classificationbse}}
\end{center}
\end{figure}

\subsubsection{Linear Correlation Coefficients}
We calculated the $P$--$F$ linear correlation coefficients, which are presented in Figure \ref{correlationbse}, and as a table in Supplemental Material. In each case, the mean value of the data was subtracted before calculating the coefficients and errors represent the standard deviation in a 10000-fold bootstrap re-sampling. The results presented in Figure \ref{correlationbse} mostly show similar behaviour as derived from the classification scheme. The analysis shows that the pre-edge intensity depends positively on broken tetrahedrality $\Delta_a$, and $\Delta_d$, whereas the opposite is the case for the post edge, ROI III. In a similar manner, the pre-edge intensity is negatively correlated with the number of hydrogen bonds, whereas the opposite is seen for the post edge ROI III.
\par
\begin{figure}[h]
\begin{center}
\includegraphics[width=9cm]{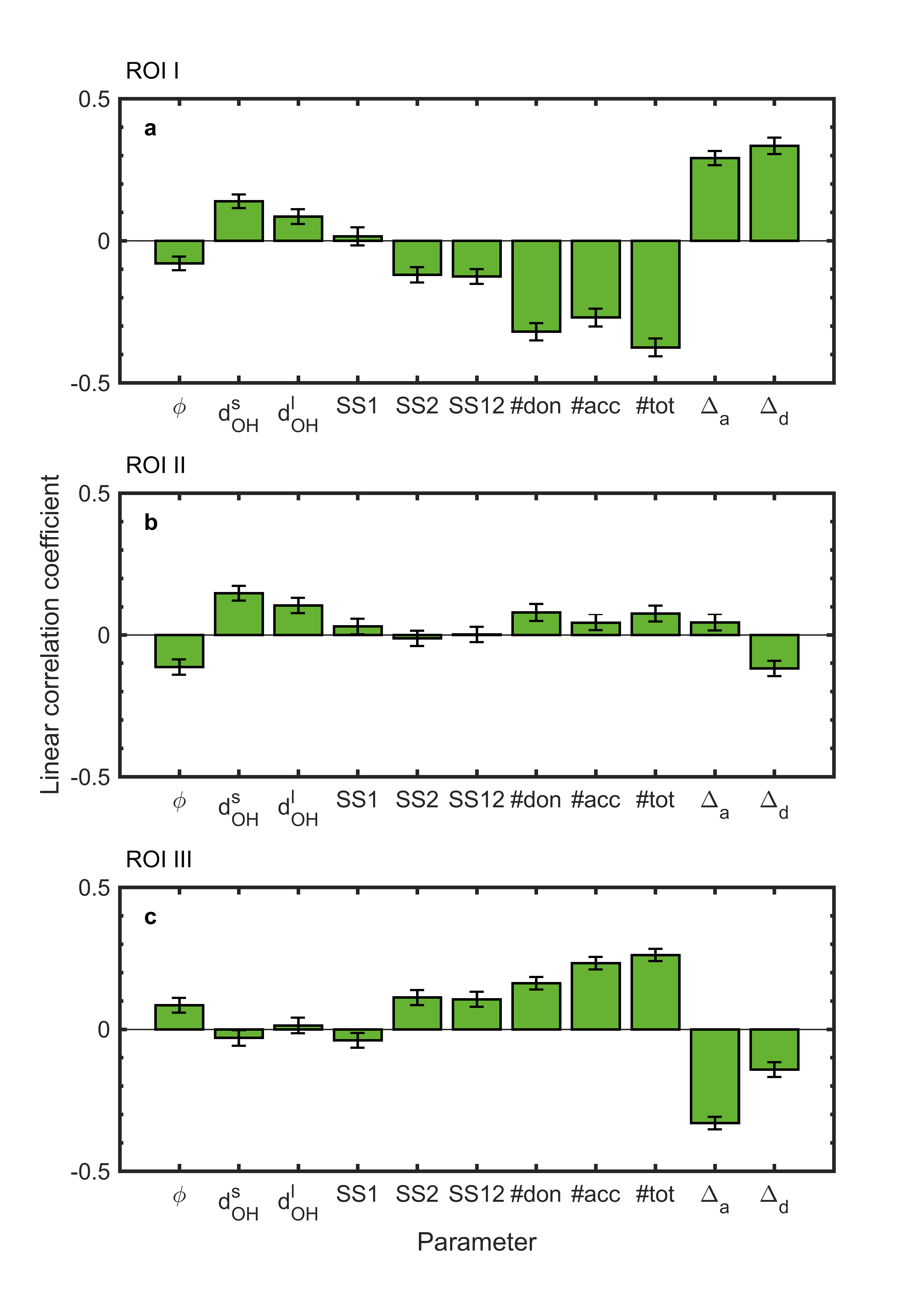}
\caption{Linear correlation coefficients between structural parameters and intensities in ROIs I, II and III.}\label{correlationbse}
\end{center}
\end{figure}

\section{Discussion}
The oxygen K-edge spectrum of water has been studied intensively during the past 15 years \cite{Wernet2004,pylkkanen2010,pylkkanen2011,sahlePNAS,watercar,tse2008,nilsson2015}, and reviewed recently \cite{anotherpetterssonnilssonreview2016}. Although the structure of water is still under debate \cite{nilsson2015}, certain spectral--structure relations have gradually become apparent and a common consensus seems to be emerging about features in the oxygen K-edge spectrum of liquid water. The pre-edge is enhanced together with the main edge in supercritical water \cite{sahlePNAS} and therefore is associated with broken hydrogen bonds. The post edge, on the other hand is enhanced in ice \cite{tse2008} and therefore indicates strong tetrahedral order and hydrogen bonding. The analysis methods presented herein succeeded in finding these spectrum--property correlations.
\par
The pre-edge (ROI I) is prominent in supercritical water \cite{sahlePNAS} and it gains spectral weight and shifts to lower energies for increasing temperature in the liquid state \cite{pylkkanen2011}. It has, therefore, been related to weakened or broken hydrogen bonds and a disturbed hydrogen bond network. However, the pre-edge also changes as a function of the approaching second solvation shell in high-pressure ice phases \cite{pylkkanen2010}, which indicates the complexity of the interrelation and spectrum--structure relations. The main-edge (ROI II) is also commonly connected with a heavily distorted and weakly hydrogen bonded water network and, for example, Tse {\it et al.} reported a decrease in spectral weight in the main-edge when comparing high- and low-density amorphous water ice \cite{tse2008}. As for the case of the pre-edge feature, changes in the main-edge have also been attributed to structural changes in which the hydrogen bonding remains the same, i.e.~Pylkk\"anen {\it et al.} reported a linear increase of the main-to-post-edge intensity ratio when comparing different ice phases of different densities but similar hydrogen bond arrangements\cite{pylkkanen2010}. The post-edge (ROI III), in contrast, is prominent in presence of tetrahedral order and a strong hydrogen-bond network. Thus, the oxygen K-edges of several ice phases exhibit increased spectral weight in the post-edge region \cite{tse2008, pylkkanen2010}. Tetrahedrality and hydrogen bonding are linked in our analysis to post-edge intensity.
\par
It is fascinating that by studying a liquid system in only one statistical ensemble, it is possible to obtain structure--spectrum (ROI intensities) correlations seen in experiments for both ice and supercritical water. This is owing to the {\it variation} of local configurations in the ensemble. Thus the analysis procedure outlined can be expected to work for cases where $\rho$ changes due to a change in experimental conditions, but the system is not pushed to completely new parts of the local phase space. For example, our liquid simulation does not probe high-density-ice-like local configurations well, because these involve reorganization of the whole second solvation shell that is very unlikely to happen in the liquid state. As a result, the main edge behavior seen in Ref. \citenum{pylkkanen2010} is not probed by our analysis.
\par
Water has been proposed to appear in two coexisting liquid phases \cite{nilsson2015}. Some MD simulations have shown these two phases \cite{takuma2014}, but these results have also been questioned \cite{overduin2015}. Studying this problem with our method would be very complicated. First, real-valued mathematical definitions for the phases are required, and the results would only consider the particular definition out of its infinitely many alternatives. Second, the structural simulations should not limit two-phase phenomena (if they exist) simply because of a limited box size, and time scale. Moreover, the spectral simulation should be able to treat such boxes. All this is beyond 64-molecule systems, and therefore we limit the scope of this work to simple, maximally local, and well-understood structural parameters, for which a general consensus in terms of spectral changes along structure has been established. We note that there is indeed a wealth of liquid systems for which phenomena related to parameters like these are the main interest of the simulation.
\par
For the applied method, the nature of the spectroscopic transition is not very relevant as long as the process defines a spectrum depending dominantly on the local configuration. Due to enormous amounts of transition lines even for one local configuration, core-level absorption, emission, and resonant inelastic scattering yield $\mathcal{R}^m$-valued functions, where $m$ in practice is the number of energy channels in the relevant energy range. For a single local configuration, on the other hand, K-edge X-ray photoelectron spectroscopy typically yields a real number (the core-electron binding energy). From the point of view of this work, the more complicated spectral signals like X-ray absorption may therefore provide more information about the sample.
\section{Conclusions}
We have presented a systematic statistical analysis of simulations to help the interpretation of core-level spectra. Our approach defines regions of interest in the spectra, based on identifiable spectral features in a simulated ensemble average. The intensity of these regions allows for a quantitative analysis and interpretation of the spectral features in terms of the underlying local structural parameters. For our analysis we studied a simulation of the oxygen K-edge spectrum of liquid water, and the known data of liquid water, supercritical water, and ice support the spectrum--structure-relationship findings. Our results show that due to statistical variation, core-level spectra can also be interpreted by using MD simulations and by studying correlations between spectral intensities and structural parameters, instead of fingerprint analysis using predefined model structures.

\begin{acknowledgements}
We are grateful to Dr. Iina Juurinen and to Dr. Antti Kettunen for related discussions. We acknowledge the ESRF for providing synchrotron radiation. Christian Henriquet is kindly thanked for technical support. Ch.J.S.~is thankful to Marco Moretti-Sala for supporting this work.
\end{acknowledgements}

\bibliography{references}

\section*{Supplemental Material}
Supplemental Material contains plots of intensities in regions I, II, and III against the structural parameters. A figure representing the alignment of the spectra for the 64-spectrum snapshots, the data of the classification scheme, and the linear correlation coefficients are also presented.

\begin{figure}[h]
\begin{center}
\includegraphics[width=8cm]{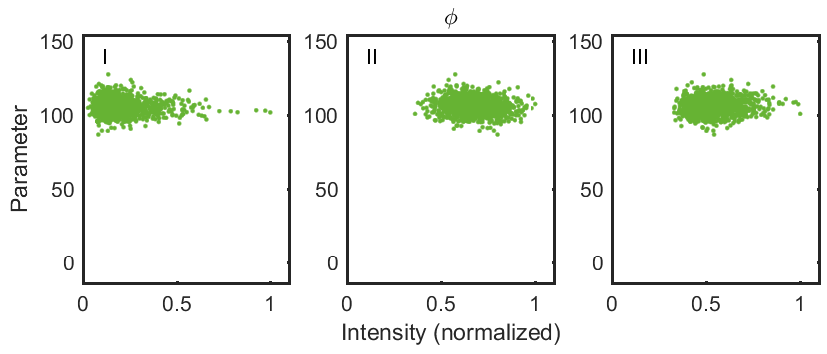}
\includegraphics[width=8cm]{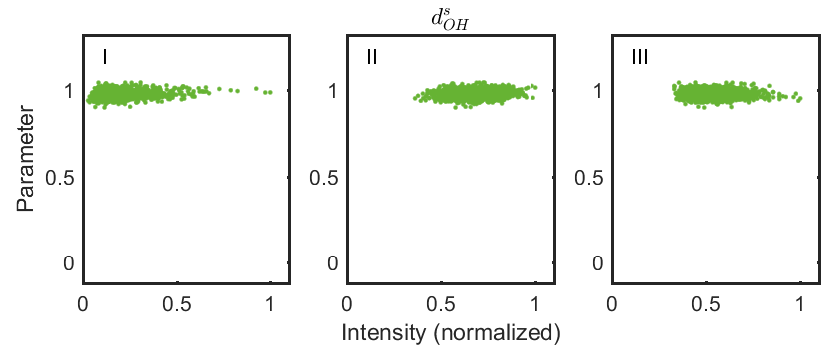}
\includegraphics[width=8cm]{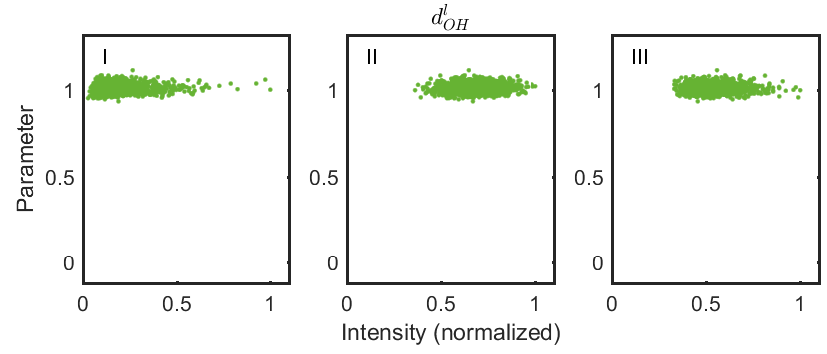}
\includegraphics[width=8cm]{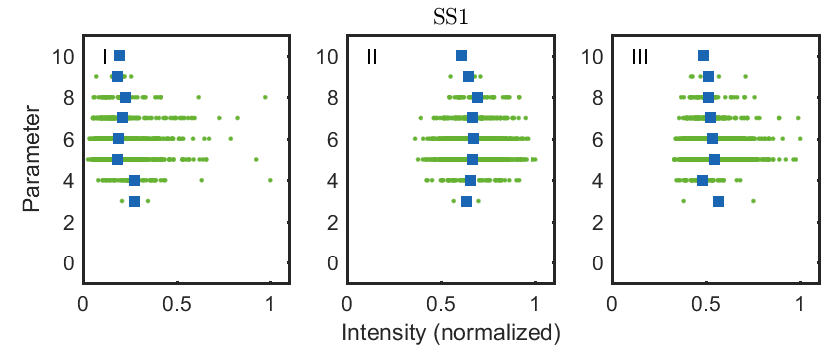}
\includegraphics[width=8cm]{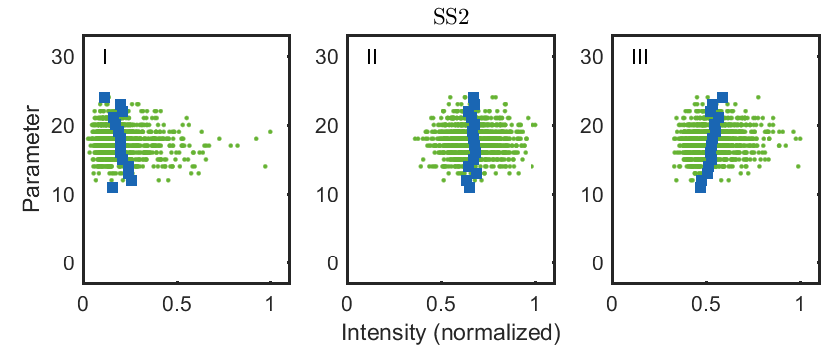}
\includegraphics[width=8cm]{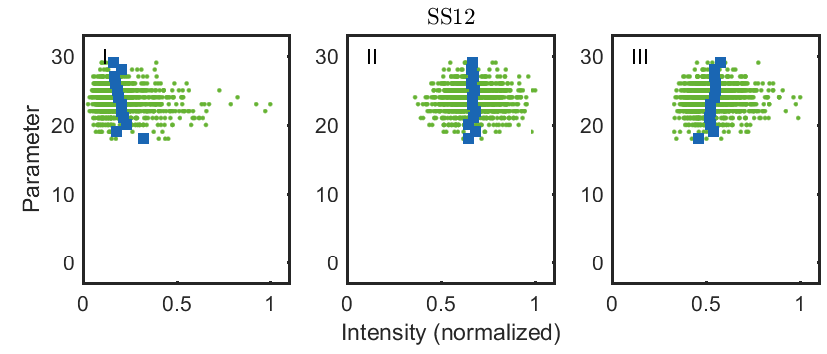}
\end{center}
\caption{\label{appfig1}The intensity data in regions I, II, and III for bonding angle, for the shorter OH bond, for the longer OH bond, and for numbers of water molecules in solvation shells. Each green marker corresponds to a local configuration in the simulation. The blue markers show the mean intensity value for the discrete-valued parameter values.}
\end{figure}

\begin{figure}[h]
\begin{center}
\includegraphics[width=8cm]{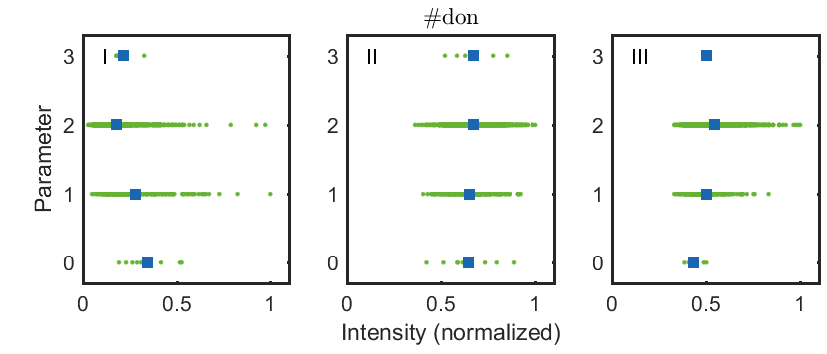}
\includegraphics[width=8cm]{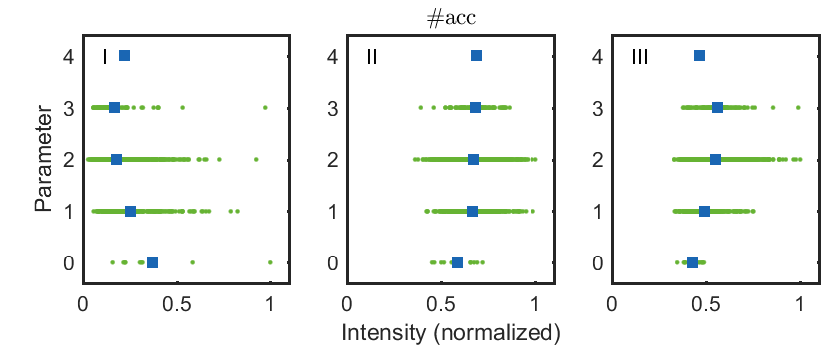}
\includegraphics[width=8cm]{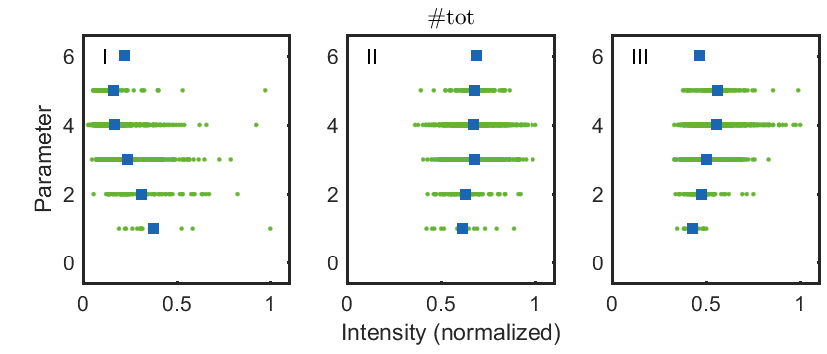}
\includegraphics[width=8cm]{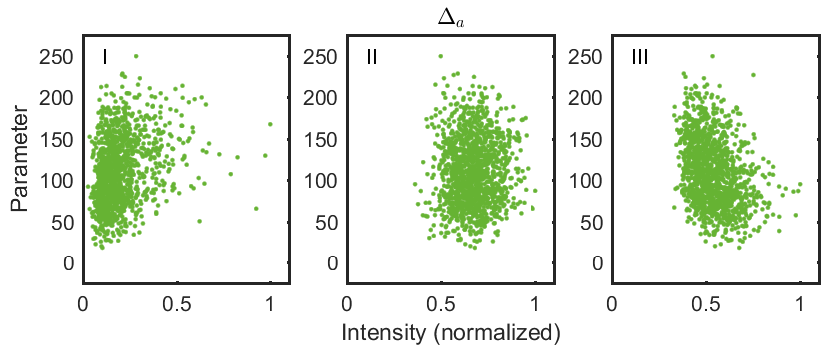}
\includegraphics[width=8cm]{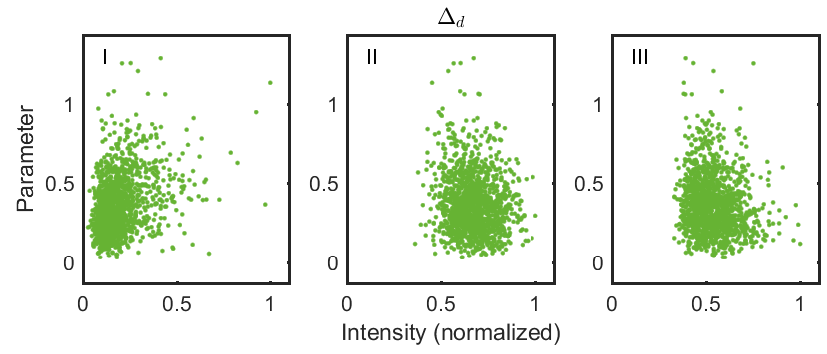}
\end{center}
\caption{\label{appfig4}The intensity data in regions I, II, and III for numbers of hydrogen bonds made by the molecule, and for deviation from the tetrahedral geometry. Each mark corresponds to a local configuration in the simulation. The blue markers show the mean intensity value for the discrete-valued parameter values.}
\end{figure}

\begin{figure}[h]
\begin{center}
\includegraphics[width=8cm]{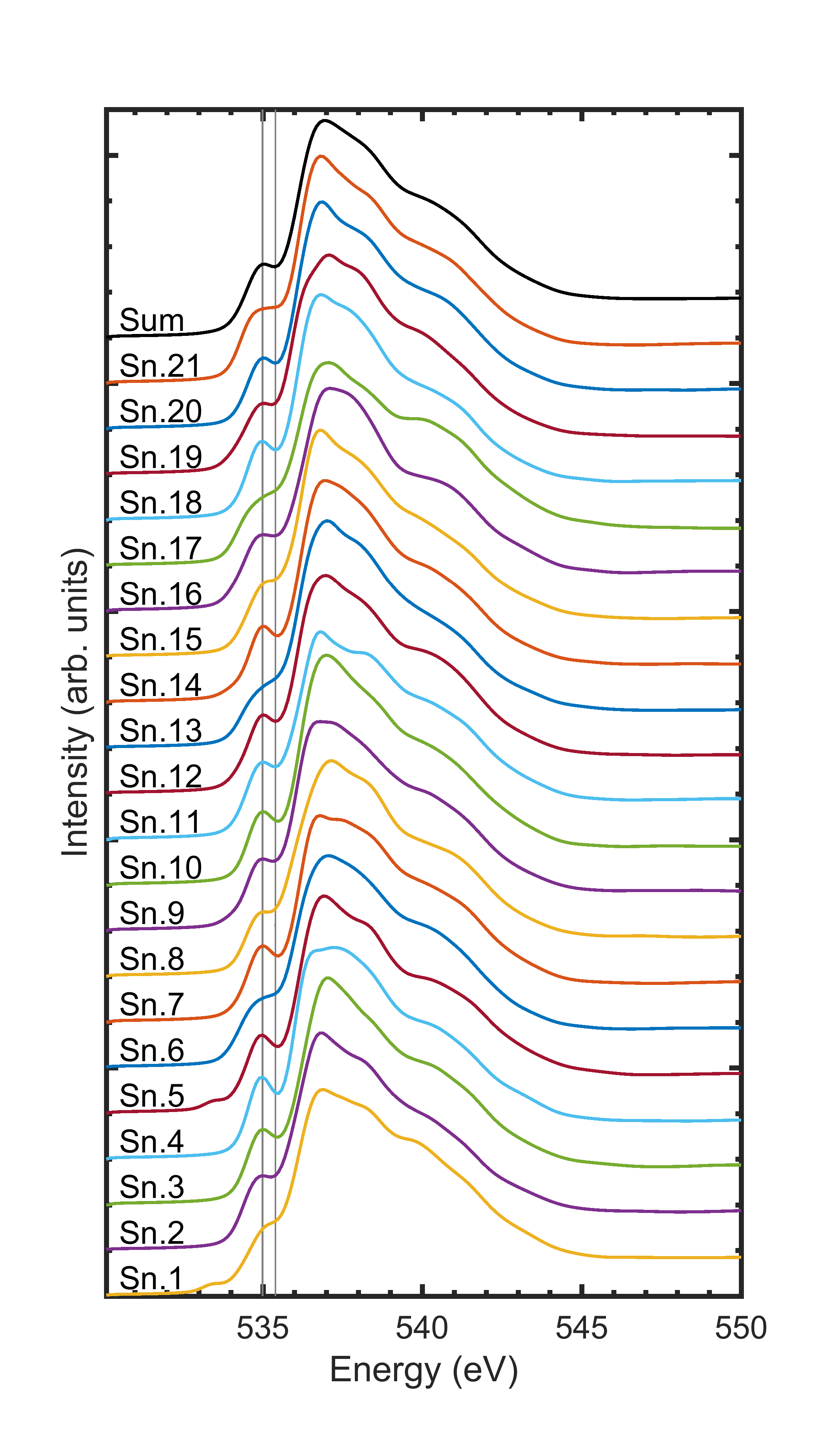}
\end{center}
\caption{\label{appfig5}Alignment of the 64-oxygen spectra from the different snapshots. The vertical lines at 535.0 eV and 535.4 eV are presented to guide the eye.}
\end{figure}

\FloatBarrier

\begin{table}[h!]
\begin{ruledtabular}
\caption{Difference between mean structural parameters for structures boosting and suppressing average ROI intensity. Positive values mean that an increase in the parameter is associated with an increase in ROI intensity. The total average is also given. \label{classificationbse2}}
\begin{tabular}{l r r r r}
\textbf{Parameter} & \textbf{I} & \textbf{II} & \textbf{III} & \textbf{ave.}\\
$\phi$ & -0.93 (0.29) & -0.92 (0.28) & 0.87 (0.28) & 105.87\\
$d_{OH}^{s}$ & 0.00 (0.00) & 0.01 (0.00) & 0.00 (0.00) & 0.98\\
$d_{OH}^{l}$ & 0.00 (0.00) & 0.00 (0.00) & 0.00 (0.00) & 1.01\\
SS1 & -0.03 (0.06) & 0.02 (0.05) & -0.05 (0.05) & 5.70\\
SS2 & -0.32 (0.12) & -0.03 (0.11) & 0.46 (0.11) & 17.86\\
SS12 & -0.36 (0.10) & -0.01 (0.10) & 0.41 (0.10) & 23.56\\
\#don & -0.21 (0.03) & 0.05 (0.02) & 0.11 (0.02) & 1.84\\
\#acc & -0.33 (0.03) & 0.05 (0.03) & 0.25 (0.03) & 1.84\\
\#tot & -0.55 (0.04) & 0.09 (0.04) & 0.36 (0.04) & 3.67\\
$\Delta_a$ & 21.29 (2.25) & 3.10 (2.24) & -22.06 (2.15) & 110.16\\
$\Delta_d$ & 0.11 (0.01) & -0.03 (0.01) & -0.05 (0.01) & 0.37\\
\end{tabular}
\end{ruledtabular}
\end{table}
\begin{table}[h!]
\caption{Linear correlation coefficients between structural parameters and intensities in ROIs I, II and III.}\label{correlationbse2}
\begin{ruledtabular}
\begin{tabular}{l r r r}
\textbf{Parameter} & \textbf{I} & \textbf{II} & \textbf{III} \\
$\phi$ & -0.08 (0.02) & -0.11 (0.03) & 0.08 (0.03) \\
$d_{OH}^{s}$ & 0.14 (0.02) & 0.15 (0.03) & -0.03 (0.03) \\
$d_{OH}^{l}$ & 0.09 (0.03) & 0.10 (0.03) & 0.01 (0.03) \\
SS1 & 0.02 (0.03) & 0.03 (0.03) & -0.04 (0.03) \\
SS2 & -0.12 (0.03) & -0.01 (0.03) & 0.11 (0.03) \\
SS12 & -0.13 (0.03) & 0.00 (0.03) & 0.11 (0.03) \\
\#don & -0.32 (0.03) & 0.08 (0.03) & 0.16 (0.02) \\
\#acc & -0.27 (0.03) & 0.04 (0.03) & 0.23 (0.02) \\
\#tot & -0.38 (0.03) & 0.08 (0.03) & 0.26 (0.02) \\
$\Delta_a$ & 0.29 (0.03) & 0.04 (0.03) & -0.33 (0.02) \\
$\Delta_d$ & 0.33 (0.03) & -0.12 (0.03) & -0.14 (0.03) \\
\end{tabular}
\end{ruledtabular}
\end{table}

\end{document}